\documentclass[preprint,12pt, a4paper]{elsarticle}

\usepackage[hyphens]{url}
\usepackage{hyperref}
\hypersetup{colorlinks=true,breaklinks=true,pdfauthor=author}
\usepackage{amssymb}
\usepackage{subcaption}

\graphicspath{ {./img/} }

\usepackage{listings}
\usepackage{color}
\definecolor{lightgray}{rgb}{.9,.9,.9}
\definecolor{darkgray}{rgb}{.4,.4,.4}
\definecolor{purple}{rgb}{0.65, 0.12, 0.82}

\newcommand*\justify{%
	\fontdimen2\font=0.4em
	\fontdimen3\font=0.2em
	\fontdimen4\font=0.1em
	\fontdimen7\font=0.1em
	\hyphenchar\font=`\-
}
\newcommand{\function}[1]{{\tt \justify{\textbf{#1}}}}

\lstdefinelanguage{JavaScript}{
  keywords={typeof, new, true, false, catch, function, return, null, catch, switch, var, if, in, while, do, else, case, break},
  keywordstyle=\color{blue}\bfseries,
  ndkeywords={class, export, boolean, throw, implements, import, this},
  ndkeywordstyle=\color{darkgray}\bfseries,
  identifierstyle=\color{black},
  sensitive=false,
  comment=[l]{//},
  morecomment=[s]{/*}{*/},
  commentstyle=\color{purple}\ttfamily,
  stringstyle=\color{red}\ttfamily,
  morestring=[b]',
  morestring=[b]"
}

\definecolor{eclipseStrings}{RGB}{176,99,99}
\definecolor{eclipseKeywords}{RGB}{4,81,165}
\definecolor{eclipseBackground}{RGB}{191,191,191}
\definecolor{numb}{RGB}{135, 108, 191}

\lstdefinelanguage{json}{
basicstyle=\linespread{0.9}\ttfamily\footnotesize,
upquote=true,
commentstyle=\color{eclipseStrings}, 
stringstyle=\color{eclipseKeywords}, 
numbers=none,
numberstyle=\tiny\color{codegray},
numbers=none,
numberstyle=\scriptsize,
stepnumber=1,
numbersep=8pt,
showstringspaces=false,
breaklines=true,
string=[s]{"}{"},
comment=[l]{:\ "},
morecomment=[l]{:"},
literate=
*{0}{{{\color{numb}0}}}{1}
{1}{{{\color{numb}1}}}{1}
{2}{{{\color{numb}2}}}{1}
{3}{{{\color{numb}3}}}{1}
{4}{{{\color{numb}4}}}{1}
{5}{{{\color{numb}5}}}{1}
{6}{{{\color{numb}6}}}{1}
{7}{{{\color{numb}7}}}{1}
{8}{{{\color{numb}8}}}{1}
{9}{{{\color{numb}9}}}{1}
}

\usepackage[noabbrev]{cleveref}

\usepackage{lineno}
\usepackage{threeparttable}
\usepackage{float}
\usepackage{longtable}
\usepackage{multirow}
\usepackage{multicol}

\usepackage{graphicx}
\usepackage{tikz}
\usepackage{pgfplots}
\usepgfplotslibrary{statistics}
\pgfplotsset{
    compat=1.18,
    xlabel style={font=\scriptsize},
    ylabel style={font=\scriptsize},
    tick label style={font=\scriptsize},
}
\usepackage{pgfplotstable}
\usepackage{rotating}

\journal{arxiv}
\begin{document}

\begin{frontmatter}
\title{\textit{LEI2JSON}: Schema-based Validation and Conversion of Livestock Event Information}

\author[1,2,5]{Mahir Habib\corref{correspondingauthor}}\ead{mhabib@csu.edu.au}
\author[1,2,5]{ Muhammad Ashad Kabir}\ead{akabir@csu.edu.au}
\author[1,2,5]{Lihong Zheng}\ead{lzheng@csu.edu.au}

\affiliation[1]{organization={School of Computing, Mathematics and Engineering, Charles Sturt University}, city={Bathurst}, state={NSW}, postcode={2795}, country={Australia}}

\affiliation[2]{organization={Gulbali Institute for Agriculture, Water and Environment, Charles Sturt University}, city={Wagga Wagga}, state={NSW}, postcode={2678}, country={Australia}}



\affiliation[5]{organization={Food Agility CRC Ltd}, city={Sydney}, state={NSW}, postcode={2000}, country={Australia}}

\cortext[correspondingauthor]{Corresponding author: Charles Sturt University, Panorama Ave, Bathurst, NSW 2795. Ph.+61263386259}%

\begin{abstract}
Livestock producers often need help in standardising (i.e., converting and validating) their livestock event data. This article introduces a novel solution, \emph{LEI2JSON} (\textbf{L}ivestock \textbf{E}vent \textbf{I}nformation \textbf{To} \textbf{JSON}). The tool is an add-on for Google Sheets, adhering to the livestock event information (LEI) schema. The core objective of LEI2JSON is to provide livestock producers with an efficient mechanism to standardise their data, leading to substantial savings in time and resources. This is achieved by building the spreadsheet template with the appropriate column headers, notes, and validation rules, converting the spreadsheet data into JSON format, and validating the output against the schema. LEI2JSON facilitates the seamless storage of livestock event information locally or on Google Drive in JSON. Additionally, we have conducted an extensive experimental evaluation to assess the effectiveness of the tool.
 
\end{abstract}


\end{frontmatter}




\begin{table}[H]
\begin{tabular}{|l|p{5cm}|p{5cm}|}
\hline
\textbf{Nr.} & \textbf{Code metadata description} & \textbf{Please fill in this column} \\
\hline
C1 & Current code version & v0.0.1 \\
\hline
C2 & Permanent link to code/repository used for this code version & \url{https://github.com/mahirgamal/LEI2JSON} \\
\hline
C3 & Code Ocean compute capsule & - \\
\hline
C4 & Legal Code License   & Apache License 2.0 \\
\hline
C5 & Code versioning system used & none \\
\hline
C6 & Software code languages, tools, and services used & HTML,Google Apps Script, JavaScript, CSS \\
\hline
C7 & Compilation requirements, operating environments \& dependencies & Google account\\
\hline
C8 & If available Link to developer documentation/manual & none \\
\hline
C9 & Support email for questions & \href{mailto:mhabib@csu.edu.au}{mhabib@csu.edu.au}\\
\hline
\end{tabular}
\caption{Code metadata}

\end{table}


\section{Motivation and significance}

Livestock management requires recording various events such as weight, movement, and vaccination. Producers depend on these records to maximise profits and produce high-quality meat. The Red Meat Advisory Council proposes a federal investment of \$12 million to improve the industry's quality, profitability, and sustainability~\citep{RedMeatAdvisoryCouncil2022RedSubmission}.

Livestock record-keeping is indispensable for producers. It facilitates inventory management, supports market analysis, and ensures traceability for vital aspects such as biosecurity, meat safety, product integrity, and market access. With these dependable measures in place, livestock producers operate with confidence and efficiency~\citep{Bowling2008REVIEW:America}.

To unlock the full value of this event information, it must be formatted to allow stakeholders to analyse, save in databases, or transmit as messages between systems. The JavaScript Object Notation (JSON) format excels at meeting these diverse requirements. Nevertheless, the challenge persists in the form of producers requiring more substantial technological expertise~\cite{Chuang2020,Kshetri,Chernbumroong2022}, a pursuit that demands a significant investment of both time and resources.

Producers have experience with spreadsheet management tools like Excel and Google Sheets. Additionally, the Organic Farmer's Business Handbook suggests the utilisation of spreadsheets to handle administrative tasks~\citep{wiswall2009organic}. Simultaneously, the National Livestock Identification System (NLIS) in Australia mandates producers to upload comma-separated values (CSV) files generated through spreadsheets or Notepad~\citep{NationalLivestockIdentificationSystem2013}.

Therefore, there arises a need for a tool that seamlessly generates JSON while integrating with spreadsheet management tools. Additionally, the produced JSON must adhere to a specific JSON schema\footnote{JSON Schema defines the structure of JSON data and validation constraints~\citep{pezoa2016foundations}.}. 

To address this need, we have leveraged Google Sheets and created a HyperText Markup Language (HTML) graphical user interface sidebar embedded within the sheet's interface. This interface allows producers to interact with the sheet to input their personal information, including name, address, email address, and property identification code (PIC), facilitating the generation of JSON. Once the JSON data is generated from the entered spreadsheet information, a validation process ensues, ensuring compliance with the livestock event information (LEI) schema standards. The outcome provided by the add-on is a JSON text that producers can conveniently copy, save, or share.

\section{Background}\label{sec:background}

TerraCipher's AgriTrakka Uploader\citep{agritrakka_uploader} facilitates the seamless publication of event data derived from CSV files to AgriTrakka connections. Producers can easily upload their Excel or CSV files directly to their AgriTrakka
connections. Select specific events from a drop-down menu, opening a standardised sheet template. This template serves as an interface for users to conveniently copy and paste their data, ensuring adherence to the standards prescribed by Integrity Systems~\citep{swain2023trakka}. AgriTrakka Uploader served as a fundamental building block for LEI2JSON.

CSV format is a widely used text format that stores tabular data. Each field is separated by commas, with records separated by a line of characters~\citep{Carvalho2015}. It is commonly used for data exchange between spreadsheet programs~\citep{Shafranovich2005} and accommodates text and numerical data~\citep{Mitlohner2016}. It is easily read with spreadsheet tools, facilitating large amounts of data sorting. Free web-based spreadsheet solutions like Google Sheets offer a user-friendly interface for managing CSV files.

Google Sheets application is an online spreadsheet editor and a component of Google Workspace~\citep{mansor2012managing}. It offers standard spreadsheet functionality, including inserting, deleting, and rearranging rows and columns~\citep{Gonzalez2010,Conner2008} and connects to other Google apps and external data sources~\citep{broman2018data}. Developers or users extend and automate Google Workspace applications, including Google Sheets, using Google Apps Script~\citep{gabet2014google}, a cloud-based scripting language and platform. Google Apps Script enables the creation of custom functions for Sheets and allows integration with other Google services, such as Calendar, Drive, and Gmail. In addition, Google Apps Script acts with Google Sheets as a grid, which works with two-dimensional arrays~\citep{ferreira2014google}.

Google Apps Script allows for the creation of sidebars~\citep{Googledeveloper2022DialogsDevelopers} on the right side of a spreadsheet, which provides additional functionality. These sidebars are designed using HTML, CSS, and JavaScript, which enable the creation of a graphical user interface with various elements such as buttons, input fields, and drop-down menus. HTML defines the structure of the sidebar's elements, while CSS provides visual styling, and JavaScript adds interactivity and handles user actions~\citep{cutler1997using,Cutler1999NewRetrieval,Mesbah2012AutomatedMaintenance}.

The LEI\footnote{\url{https://github.com/mahirgamal/LEI-schema}} schema is designed for sharing information about livestock events. It encompasses 34 different livestock events, ranging from \textit{calving} to \textit{death}. It employs a JSON format to define the structure, content, and validation rules. The primary objective is to improve data quality and ensure consistency among systems and stakeholders, encompassing a comprehensive set of 325 properties related to key components such as event dates, owner details, source information, and events. The LEI is actively used by producers, processors, regulators, researchers, and others to record and share information about livestock events.

Our tool, LEI2JSON, aims to standardise Google Sheets data according to the LEI schema. Producers input their personal information in the sidebar, choose an event schema file, build a spreadsheet template, generate JSON text, and validate it against the LEI schema. The output is a JSON text that can be saved locally, stored on Google Drive, or copied for various purposes, including data sharing.

\section{Related work}\label{sec:literature-review}

The versatility of Google Sheets has been explored in various domains, including education and business. In academic management, \citet{Mansor2012} highlighted the potential of cloud-based platforms, such as Google Sheets, to handle student records, such as grades and attendance.

In the realm of data management and integration, RDF123~\citet{Han2008} translates spreadsheet data into RDF based on user-defined mappings, similar to our approach. However, our tool extends this by creating a LEI schema-based spreadsheet template, generating JSON data, and validating it against the schema. Keemei~\citet{Rideout2016} is a Google Sheets add-on that validates bioinformatic file formats within the spreadsheet, enhancing data integrity and collaboration in bioinformatics. OPEnS Hub~\citep{DeBell2019} collects real-time data from various sensors and is applicable beyond environmental monitoring. It uses Google Sheets via PushingBox API and Google Apps Script, demonstrating the potential of real-time data collection. In library and information science, MatchMarc~\citep{Suranofsky2019} uses Google Sheets to validate bibliographic records while offering a user interface for customisation. Google Sheets has also been used in logistics~\citep{Rahman2021}, synthetic biology~\citep{nguyen2020intent,Nguyen2022}, and other fields, showcasing its versatility.

Considering these diverse applications and thoroughly analysing various studies, it becomes evident that Google Sheets' versatility and capabilities make it a valuable solution across diverse fields. Our research leverages these strengths to benefit the livestock industry, offering substantial advantages.

To provide further context,~\Cref{tab:json_tools} compares LEI2JSON with various add-ons available on the Google Workspace Marketplace. These add-ons include ImportJSON~\citep{importjson}, which allows users to import JSON data into Google Sheets, Sheet to JSON~\cite{sheet_to_json}, which simplifies the process of transforming Google Sheets into JSON files, Sheets\texttrademark~to JSON~\citep{sheet_tm_to_json}, which facilitates the conversion of JSON into named ranges within a Google Sheet, Export Sheet Data~\citep{export_sheet_data} which enables users to export their sheets as XML or JSON, and Data Connector~\citep{data_connector} which extracts data from various APIs into Google Sheets.

\begin{table}[!ht]
\caption{A comparative analysis of LEI2JSON with existing tools for JSON data integration}
\label{tab:json_tools}
\centering
\resizebox{\textwidth}{!}{%
\begin{threeparttable}
\begin{tabular}{p{4cm}p{2cm}p{2cm}p{2cm}p{2cm}p{2cm}p{2cm}p{2cm}}
\hline
Tool name & Schema to spreadsheet & Column flexible order & Generate JSON & Validation & Copy & Download & Save to~drive \\
\hline

ImportJSON~\citep{importjson} & $\times$ & $\times$ & $\times$ & $\times$ & $\times$ & $\times$ & $\times$ \\

Sheet~to~JSON~\citep{sheet_to_json} & $\times$ & $\times$ & \checkmark & $\times$ & \checkmark & $\times$ & \checkmark \\

Sheets\texttrademark~to~JSON~\citep{sheet_tm_to_json} & $\times$ & $\times$ & \checkmark & \checkmark & $\times$ & $\times$ & $\times$ \\

Export~Sheet~Data~\citep{export_sheet_data} & $\times$ & $\times$ & \checkmark & $\times$ & \checkmark & $\times$ & \checkmark \\

Data~Connector~\citep{data_connector} & $\times$ & $\times$ & \checkmark & $\times$ & \checkmark & $\times$ & \checkmark \\

LEI2JSON & \checkmark & \checkmark & \checkmark & \checkmark & \checkmark & \checkmark & \checkmark \\
\hline
\end{tabular}

\begin{tablenotes}

    \item \textbf{Schema to spreadsheet:} denotes converting a JSON schema into spreadsheet columns, validation rules, and formats.
    \item \textbf{Columns flexible order:} implies the columns can be rearranged, which does not affect the generation process.
    \item \textbf{Generate JSON:} is about generating JSON data from the spreadsheet.
    \item \textbf{Validate:} refers to validating JSON data against the JSON schema.
    \item \textbf{Copy:} allows users to copy the JSON data to the clipboard.
    \item \textbf{Download:} lets users save JSON data to their device.
    \item \textbf{Save to drive:} offers the option of storing the JSON in Google Drive.

\end{tablenotes}
\end{threeparttable}
}
\end{table}

Our innovative tool distinguishes itself from previous works in several crucial aspects. First, our add-on focuses on the specialised domain of livestock events and utilises a predefined JSON schema, such as the LEI schema, to construct a comprehensive spreadsheet template with built-in validation rules and formats. This approach facilitates both the generation and validation of JSON data. Second, our add-on operates bidirectionally: it creates column headers from a JSON schema and generates JSON data from a spreadsheet. Third, we have incorporated a user-friendly sidebar interface for selecting schemas, generating and validating JSON data, and saving or copying the output. Fourth, our add-on is primarily designed for farmers as the main user group, with the primary goal of simplifying their workflow in managing and reporting livestock events using Google Sheets. Lastly, there is no requirement for column ordering, allowing farmers to freely rearrange columns without affecting the generated JSON data.

\section{LEI2JSON}\label{sec:software-description}
\subsection{Architecture}\label{sec:softwarearchitecture}

\begin{figure}[!ht]
\centering
\includegraphics[width=\textwidth,height=\textheight,keepaspectratio]{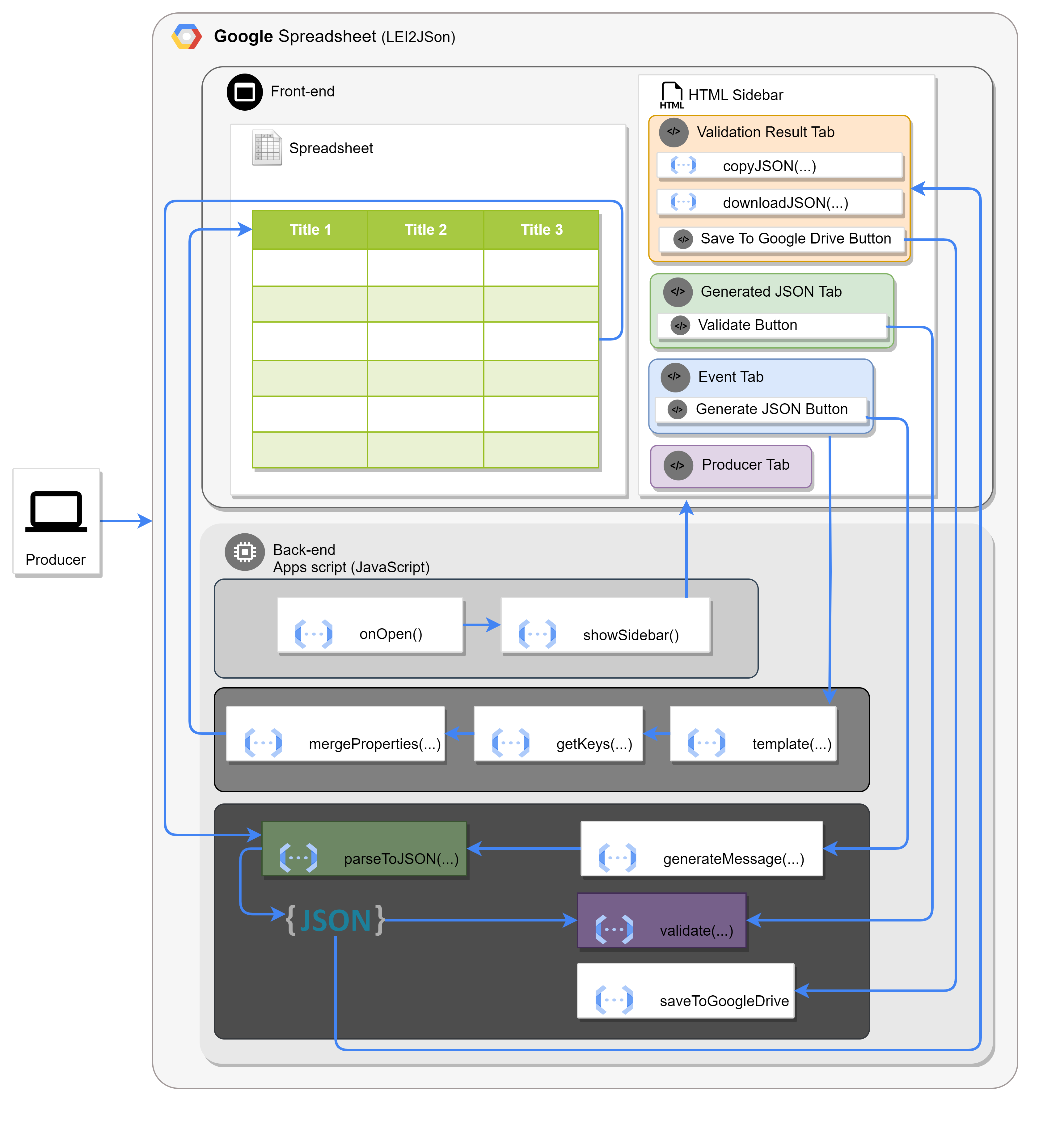}
\caption{LEI2JSON architecture}
\label{fig:publeishchitecture}
\end{figure}

The architecture of the LEI2JSON, as shown in~\Cref{fig:publeishchitecture}, comprises two main components: the front end and the back end. The front end is a Google Sheets equipped with a sidebar for interacting with the tool's functions. Additionally, it employs HTML, CSS, and JavaScript. HTML defines the page structure, CSS styles it and JavaScript adds interactivity. The back end consists of Apps Script functions responsible for extracting, generating, validating, and saving JSON data from Google Sheets.

The back end operates through three layers, collaborating to transform spreadsheet data into JSON using interconnected functions. The first layer introduces a new Google Sheets menu, LEI2JSON, with a submenu item labelled \textit{Generate JSON Message}.

The second layer's role is to extract essential data from the selected LEI JSON schema event file. This layer consists of functions such as \function{buildTemplate}, \function{getKeys}, and \function{mergeProperties}, which work together to retrieve the values of the \textit{displayName} attribute from the schema. These values serve as column headers in the first row of Google Sheets. Simultaneously, they extract the \textit{description} corresponding to each column header value and append it as a cell note to the respective header. Furthermore, this layer enforces data-type specifications for each column and implements data validation rules, including dropdown lists, which are created when there is an enumeration present within the property definition in the schema. Subsequently, this layer organises a new JSON structure for downstream processing in the next layer. Ultimately, this layer plays a central role in formulating the spreadsheet template.

The third layer generates, validates, and saves the final JSON output. This layer consists of the functions \function{generateMessage}, \function{parseToJson}, \function{validate}, and \function{saveToGoogleDrive}. It uses the new JSON structure from the previous layer as a template for each row, parsing each row into a JSON object and accumulating all the generated JSON objects into a single JSON array. A validation process is then performed to ensure the accuracy and integrity of the generated JSON array data, examining the data to determine its adherence to predefined rules. The final task for this layer is to save the JSON array, which can be done according to the user's choice: they can copy it to the clipboard, download it to their machine, or save it to Google Drive.

\subsection{User interface}\label{sec:publeish}
\begin{sidewaysfigure}[!htbp]
\centering
\includegraphics[width=\textwidth,height=\textheight,keepaspectratio]{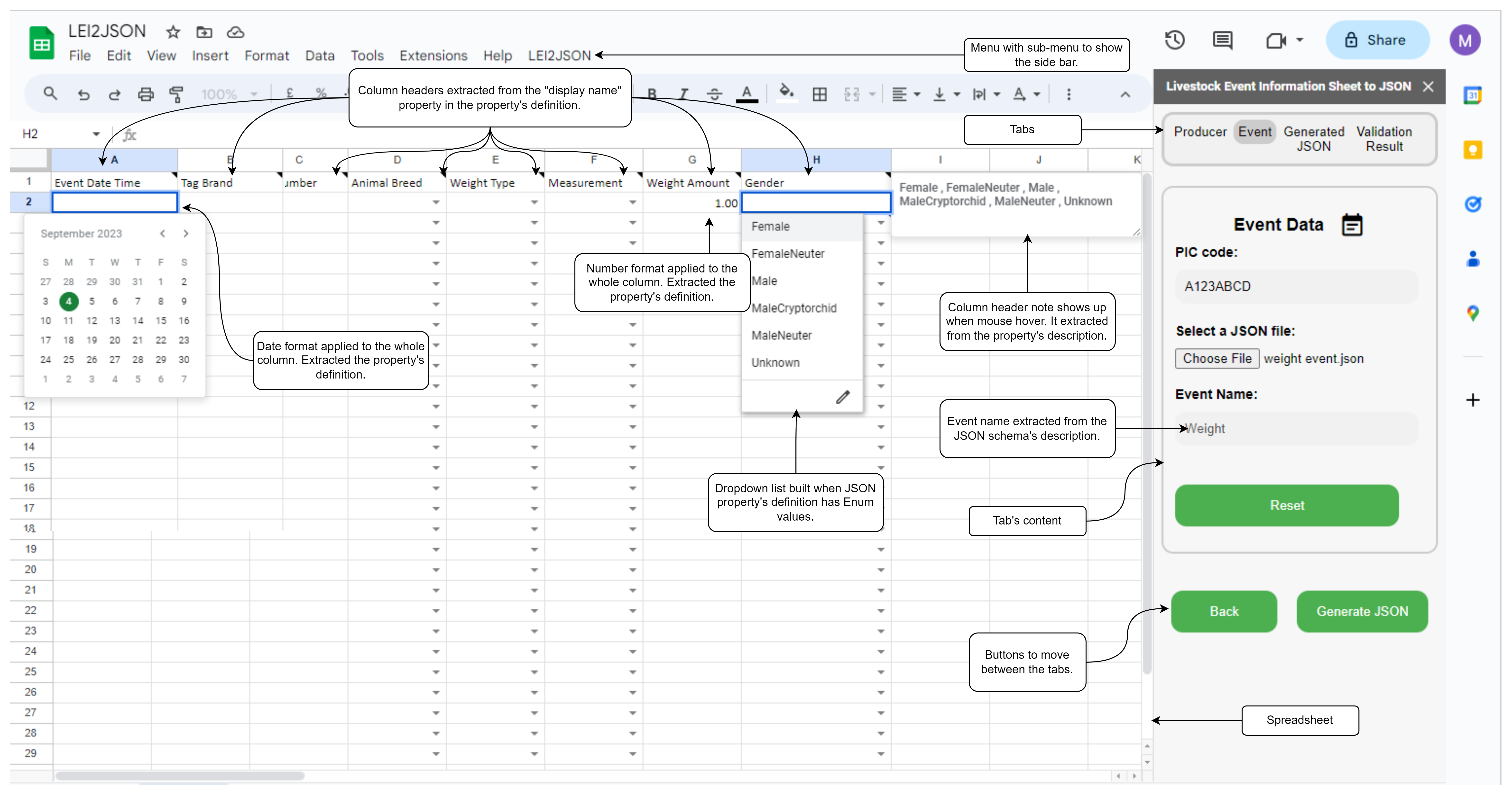}
\caption{Spreadsheet template featuring column headers, notes, validation rules, and formatting derived from JSON schema}
\label{fig:buildTemplate}
\end{sidewaysfigure}

\Cref{fig:buildTemplate} shows the implementation of the HTML sidebar and the constructed spreadsheet template. The sidebar allows the producer to select an event file to build the spreadsheet template, provide personal information, view the generated JSON before validation, perform the validation, view the validation result, and save the final generated JSON data. 

The HTML sidebar has four tabs: \textit{Producer}, \textit{Event}, \textit{Generated JSON}, and \textit{Validation Result}. Producers switch between tabs by clicking on the \textit{Next} and \textit{Back} buttons. The label on the Next button dynamically changes depending on the task it will perform when clicked.

\begin{sidewaysfigure}[!htbp]
\centering
\begin{subfigure}{0.24\textwidth}
\centering
\includegraphics[width=\textwidth,height=\textheight,keepaspectratio]{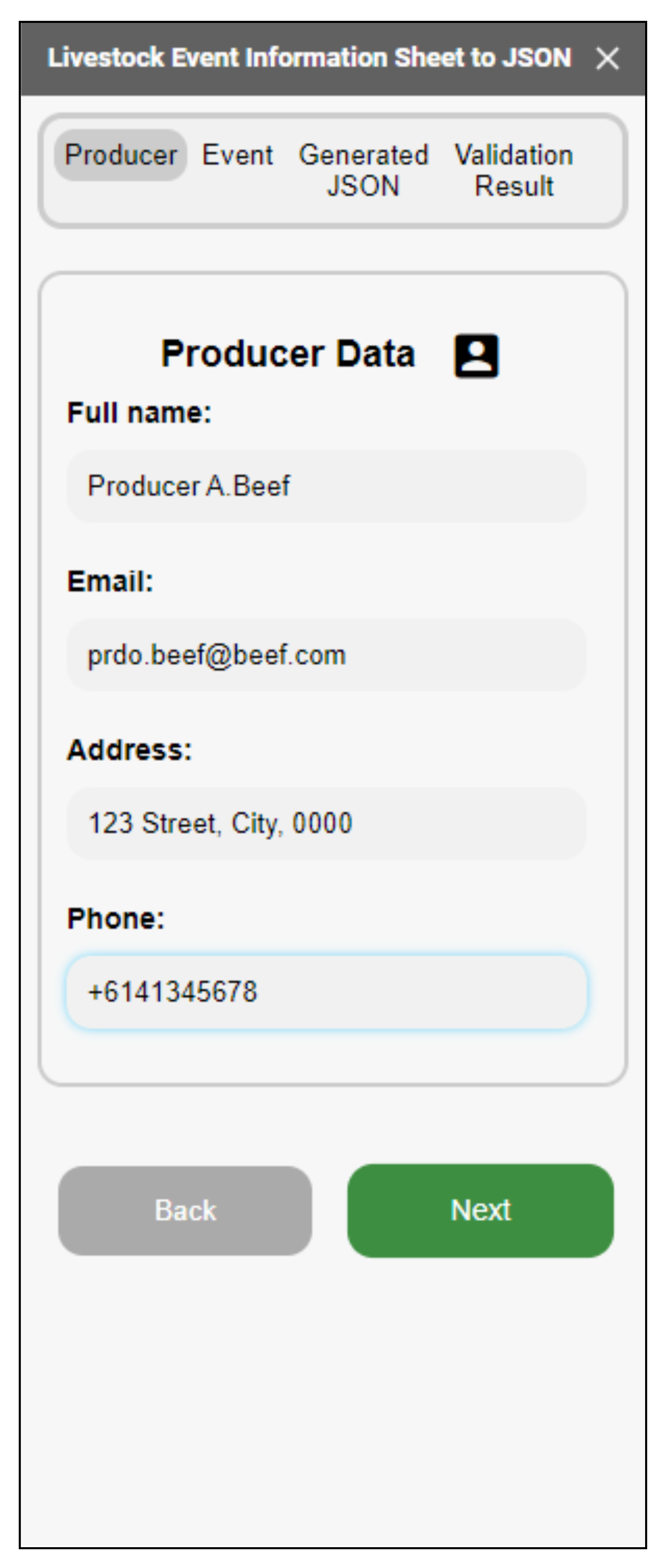} 
\caption{Producer personal data tab}
\label{fig:producer}
\end{subfigure}
\begin{subfigure}{0.24\textwidth}
\centering
\includegraphics[width=\textwidth,height=\textheight,keepaspectratio]{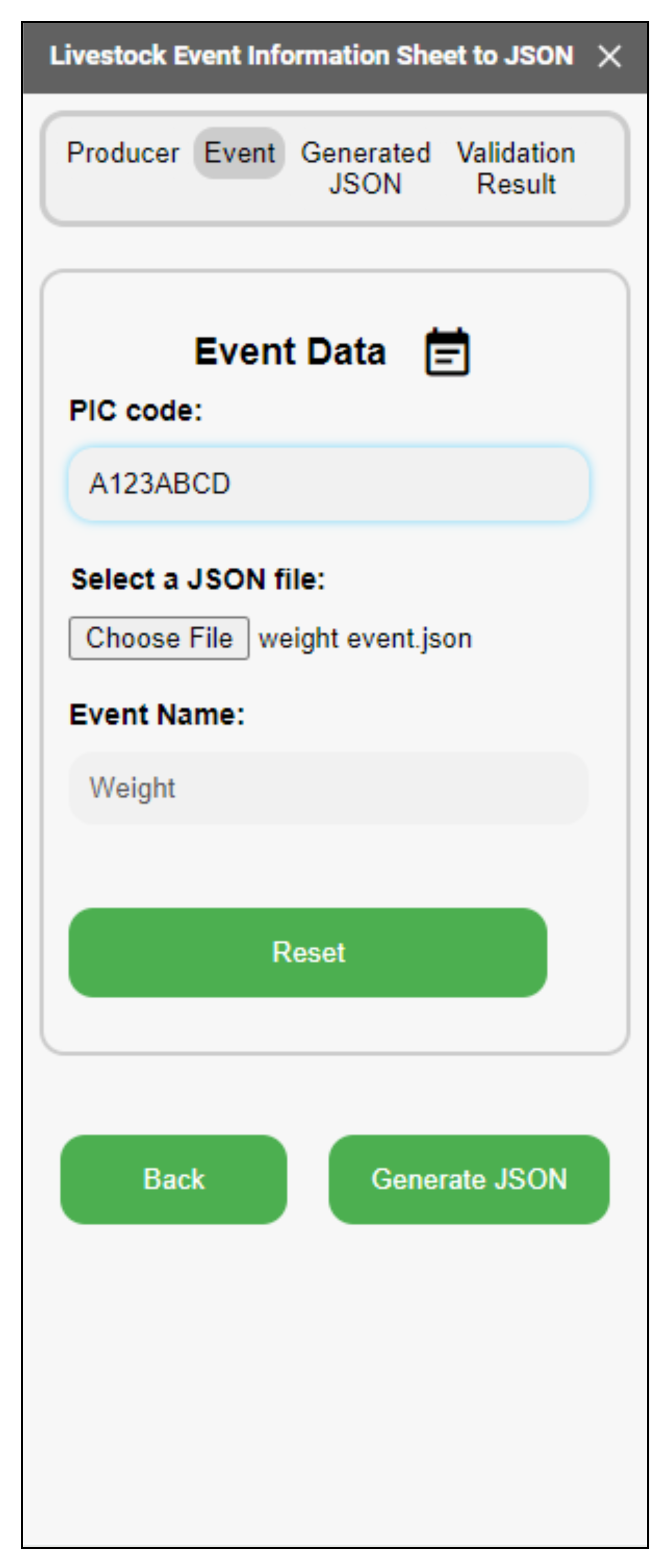} 
\caption{Producer PIC and Event data tab}
\label{fig:event}
\end{subfigure}
\begin{subfigure}{0.24\textwidth}
\centering
\includegraphics[width=\textwidth,height=\textheight,keepaspectratio]{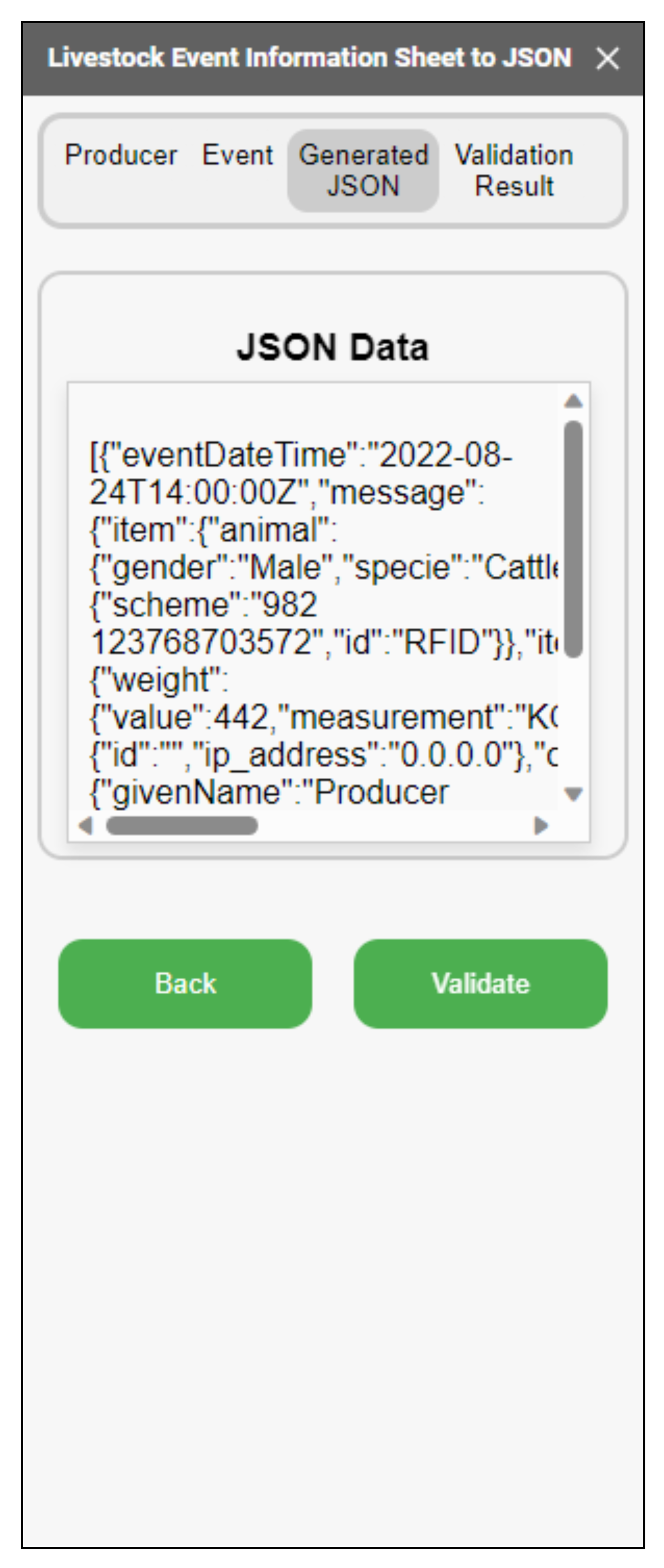} 
\caption{Producer PIC and Event data tab}
\label{fig:generated_json}
\end{subfigure}
\begin{subfigure}{0.24\textwidth}
\centering
\includegraphics[width=\textwidth,height=\textheight,keepaspectratio]{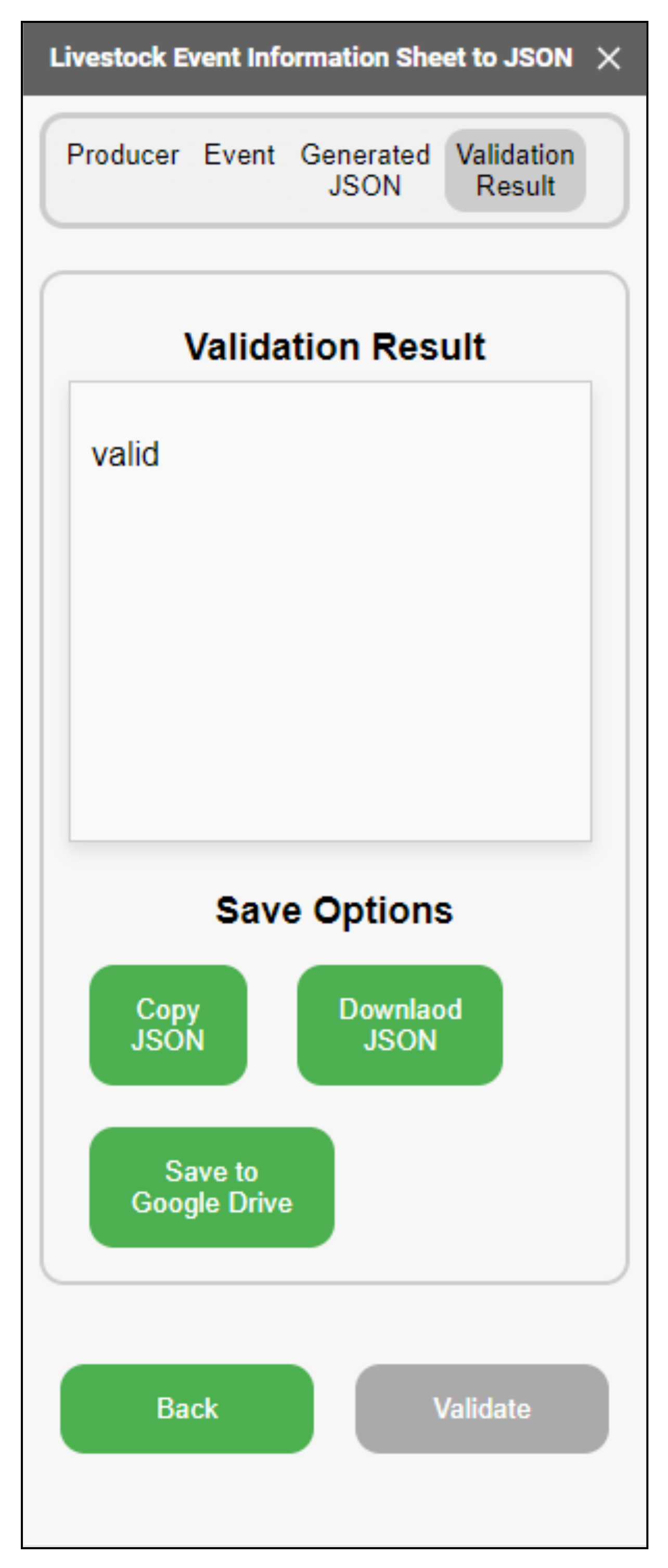} 
\caption{Producer PIC and Event data tab}
\label{fig:validatio_result}
\end{subfigure}
\caption{HTML sidebar tabs content}
\label{fig:sidebar_tabs}
\end{sidewaysfigure}

In~\Cref{fig:producer}, various input fields are available for the producer to enter their personal information. \Cref{tab:producer-data} provides descriptions of these data fields.

\begin{table}[!ht]
\caption{Producer data description }
\label{tab:producer-data}
\centering
\begin{tabular}{lp{10.5cm}}

 \hline
Field & Description \\
\hline

Full name & Producer's complete name.\\

Email & Emails have two parts: the username and the domain name separated by `@'.\\

Address & The property (i.e., farm or business) address.\\

Phone & A sequence of digits assigned to a telephone line or mobile phone.\\
\hline

\end{tabular}

\end{table}

\Cref{fig:event} allows producers to enter the PIC code and select a JSON file. When selecting the event file, the spreadsheet's first row of relevant columns automatically populates with headers, notes, validation rules, and data type specifications. The event name is also extracted and displayed. The \textit{Reset} button clears the file, event name fields, and spreadsheet data. Producers use the \textit{Generate~JSON} button to create a JSON message, as shown in~\Cref{fig:generated_json}.

The \textit{Validate} button in~\Cref{fig:generated_json} opens the \textit{Validation Result} tab, showing the validation result in a scrollable container, as shown in~\Cref{fig:validatio_result}. This tab has three buttons: \textit{Copy~JSON}, \textit{Download~JSON}, and \textit{Save~to~Google~Drive}. Clicking these buttons enables copying of the JSON data to the clipboard, downloading the JSON file, and storing the data on Google Drive, respectively.

\subsection{Functionalities}\label{sec:functionalities}

LEI2JSON offers four key functionalities: building a spreadsheet template, converting spreadsheet data to JSON, validating JSON against the schema, and saving the JSON data.

The \function{buildTemplate} function is designed to create a spreadsheet template. This function calls two functions: \function{getKeys} and \function{mergeProperties}. The \function{getKeys} function is responsible for extracting values from the schema's description for event name, displayName for column headers, type, and format for column data types and validation rules, respectively, and property description for column header notes. The \function{mergeProperties} function generates a new JSON structure to be used as a template for JSON generation for each row.

\begin{figure}[!ht]
\centering
\begin{subfigure}{\textwidth}
\centering
\includegraphics[width=\textwidth,height=\textheight,keepaspectratio]{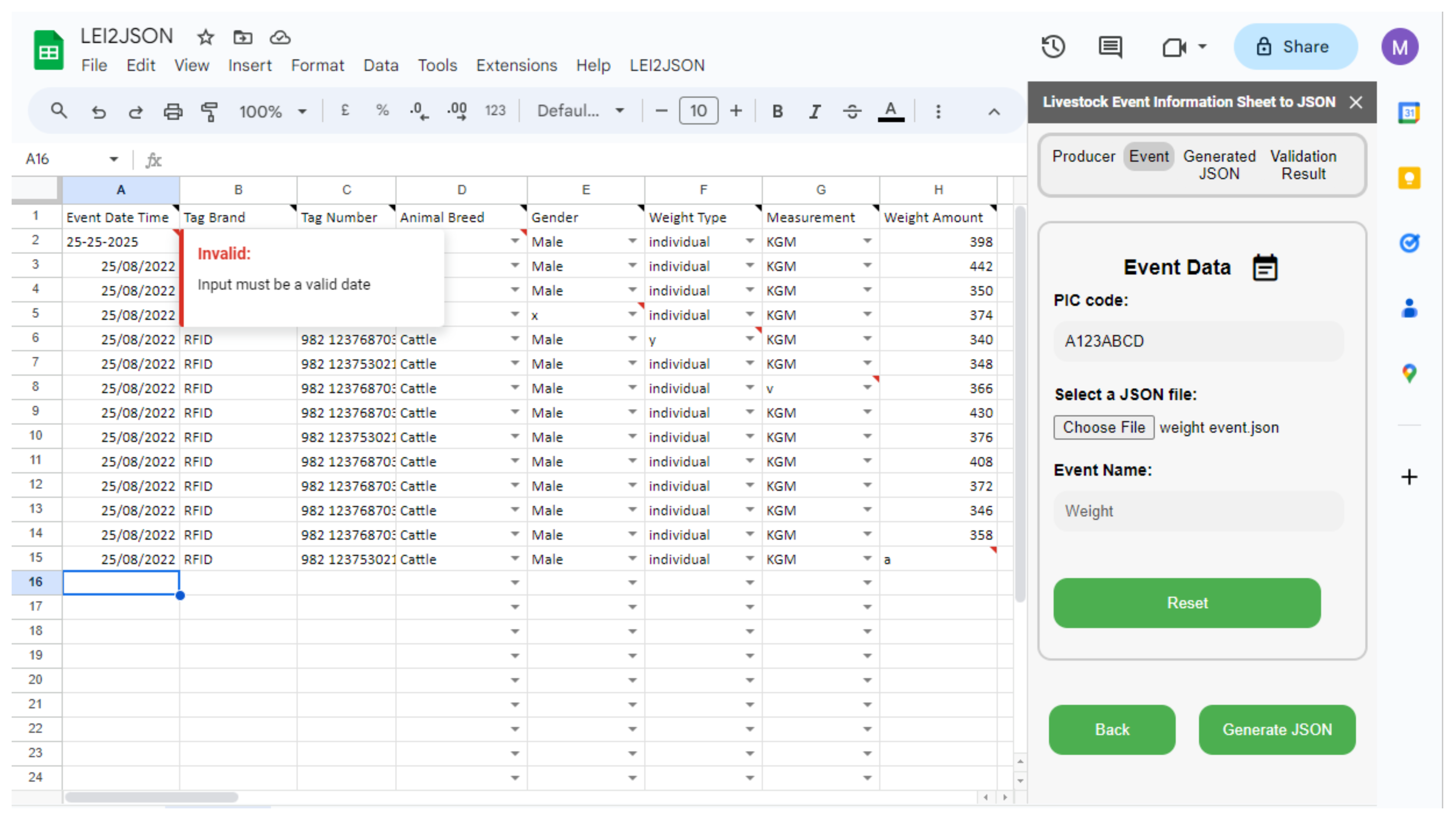}
\caption{}
\label{fig:data_checking_a}
\end{subfigure}
\begin{subfigure}{\textwidth}
\centering
\includegraphics[width=\textwidth,height=\textheight,keepaspectratio]{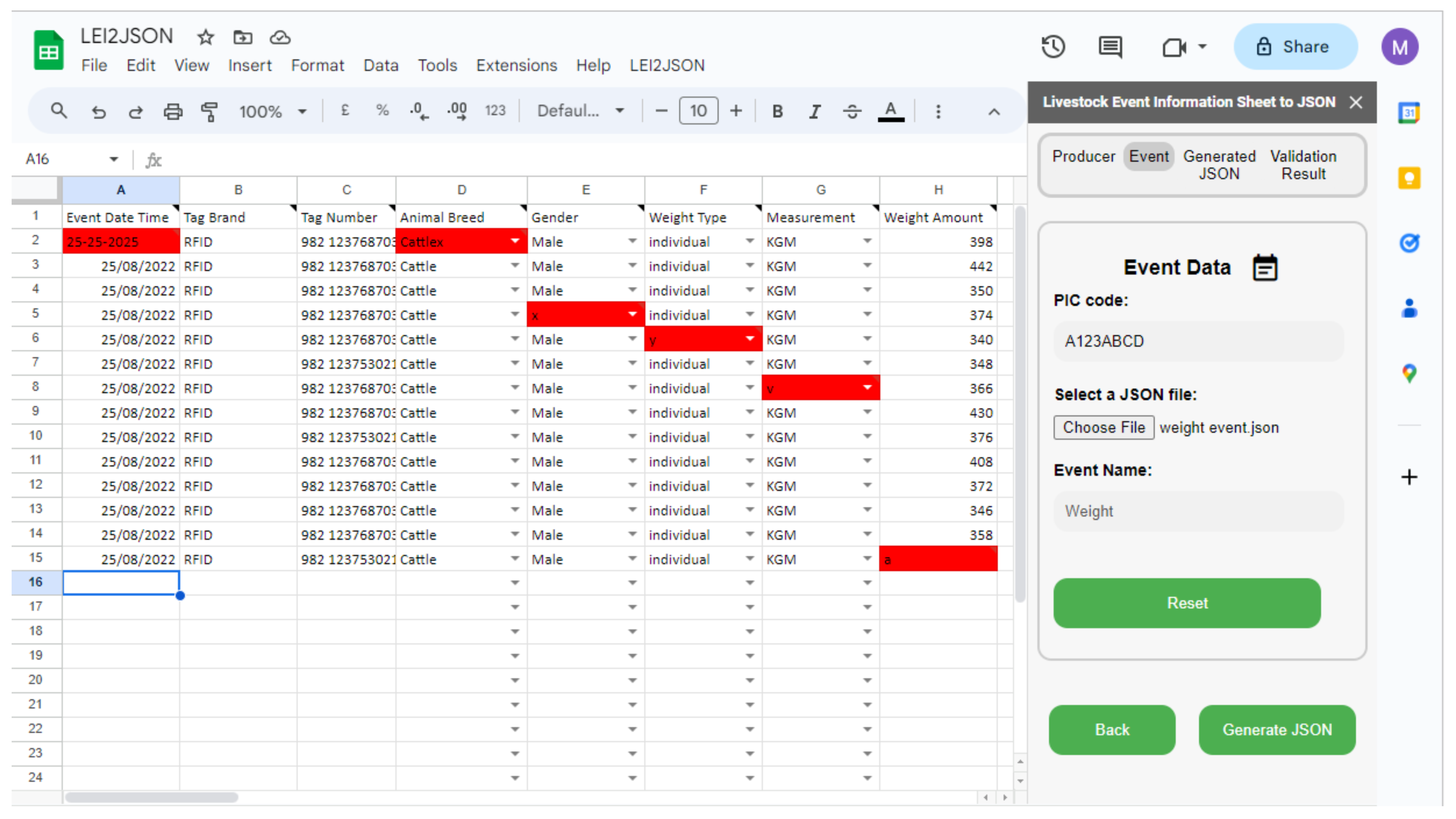}
\caption{}
\label{fig:data_checking_b}
\end{subfigure}
\caption{Verification of spreadsheet data against pre-defined column-specific validation and formatting parameters occurs when non-conforming data types or formats are entered}
\label{fig:data_checking}
\end{figure}

The \function{generateMessage} function is designed to generate JSON from the spreadsheet data. It utilises the producer's information, the event name, and the new JSON structure. Additionally, it performs real-time spreadsheet data validation based on validation rules and data types. Google Sheets automatically generates error notes in cells with errors, as shown in~\Cref{fig:data_checking_a}. Furthermore, if producers ignore the errors and attempt to generate JSON, it will highlight the cell in \textit{red}, as shown in~\Cref{fig:data_checking_b}, and prevent further tasks from being performed.

\begin{figure}[!ht]
\centering
\includegraphics[width=\textwidth,height=\textheight,keepaspectratio]{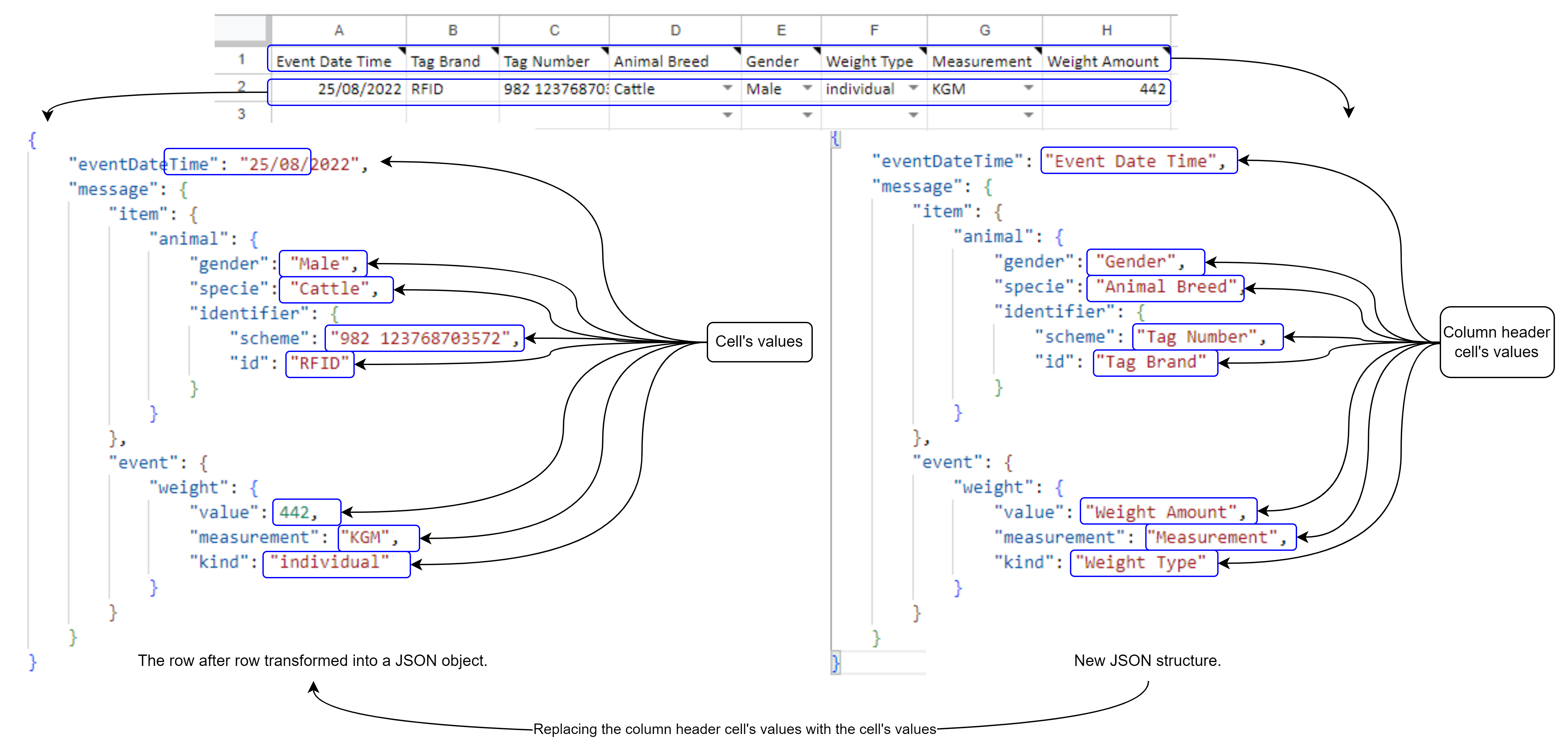}
\caption{New JSON structure used for transforming row's data to JSON by replacing column header cell's values with the header corresponding cell's values}
\label{fig:newJSONstructure}
\end{figure}

The \function{generateMessage} function calls the \function{parseToJSON} function in case invalid values are not found to parse the spreadsheet rows into JSON objects. As shown in~\Cref{fig:newJSONstructure}, this process transforms each row into a JSON object by replacing the column header cell values with the corresponding cell values within the new JSON structure. Consequently, all of these JSON objects are grouped into a JSON array.

The \function{validate} function is designed to validate the JSON array against an LEI schema by sending a POST request to a REST API hosted on AWS with the JSON message as the payload. The API returns the response code with a detailed message indicating the result of the validation process.

The functions \function{copyJSON}, \function{downloadJSON}, and \function{saveToGoogleDrive} are designed to save the JSON array. Specifically, \function{copyJSON} copies it to the clipboard, \function{downloadJSON} enables it to be downloaded as a file, and \function{saveToGoogleDrive} stores it as a file on Google Drive. It is worth noting that \function{copyJSON} and \function{downloadJSON} run within HTML, on the front end, while \function{saveToGoogleDrive} operates in Apps Script, on the back end.

\section{Experimental evaluation}

To evaluate the efficiency of LEI2JSON, we focused on its key functionalities: template creation, JSON generation, and validation. We conducted functional efficiency tests to assess how well the LEI2JSON tool performs these tasks, particularly in terms of speed. \Cref{tab:testing_machine} presents information about the testing machine and its configuration.

\begin{table}[!htbp]
\caption{Testing machine specifications}
\label{tab:testing_machine}
\centering
\begin{tabular}{ll}
 \hline
Attribute & Specification \\
 \hline
Device Type & Laptop\\
Processor & 11th Gen Intel Core i7-1165G7 @ 2.80GHz \\
Installed RAM & 16.0 GB (15.7 GB usable) \\
System type & 64-bit, x64-based \\
Operating system & Windows 11 Home \\
Operating system version & 22H2 \\
Validator API server & Apache Tomcat V9.0 \\
\hline
\end{tabular}    
\end{table}

We examine the efficiency of the tool to ensure flawless functionality, impeccable data conversion, and saving. This involved calculating the standard deviation of execution times for different functions, including \function{buildTemplate}, \function{parseToJSON}, and \function{validate}, to create a template with columns, apply all required data validation and formatting, generate, and validate JSON related to a livestock \textit{weight} event.

\pgfplotstableread[col sep=space]{
properties	data1	data2	data3	data4	data5	data6	data7	data8	data9	data10
5	205	193	196	197	191	207	197	218	205	200
10	283	278	261	269	281	270	270	304	287	253
15	339	355	353	342	337	380	309	311	350	337
20	362	401	350	385	380	341	363	363	342	393
25	447	421	434	401	459	435	402	459	403	447
}\buildTemplate

\pgfplotstableread[col sep=space]{
events  data1   data2   data3   data4   data5   data6   data7   data8   data9   data10
1000	1823	1733	1802	1766	1899	1860	1780	1724	1695	1964
2000	2948	2663	2674	2652	2799	2907	2708	2873	3104	2761
3000	4537	4388	4664	4590	4649	4154	4810	4296	4459	4495
4000	4820	4575	4935	4865	5110	5009	5155	4438	5091	4816
5000	7066	6798	6800	7088	6993	6983	6798	7099	6554	6703
6000	8088	7964	7980	7472	7513	7668	7936	7912	7645	7903
7000	9900	9677	9625	9292	9880	9478	9694	9779	9957	9240
8000	10273	9981	10173	10396	10148	10300	10085	10507	10185	10145
9000	12041	11723	11807	11732	11640	11704	11920	11908	11774	11747
10000	14306	14120	13689	13988	13874	14090	13785	14180	14111	13958
}\parseToJSON

\pgfplotstableread[col sep=space]{
events  data1   data2   data3   data4   data5   data6   data7   data8   data9   data10
1000	261	262	268	269	267	262	272	261	260	273
2000	512	521	516	550	543	525	539	545	555	550
3000	854	844	816	795	869	812	840	794	805	828
4000	1059	1080	1050	1069	1087	1063	1062	1075	1082	1095
5000	1321	1347	1299	1321	1290	1304	1309	1350	1312	1314
6000	1627	1641	1583	1642	1630	1592	1593	1558	1628	1642
7000	1870	1859	1931	1844	1891	1812	1830	1837	1813	1839
8000	2613	2545	2692	2710	2632	2585	2620	2698	2617	2754
9000	2686	2649	2818	2776	2688	2638	2717	2691	2819	2697
10000	2858	2956	2911	3064	2985	3092	3079	2847	3034	2847
}\validate

\pgfplotstablecreatecol[
    create col/expr={
        (\thisrow{data1} + \thisrow{data2} + \thisrow{data3} + \thisrow{data4} + \thisrow{data5} + \thisrow{data6} + \thisrow{data7} + \thisrow{data8} + \thisrow{data9} + \thisrow{data10})/10
    }
]{buildTemplate_mean}\buildTemplate

\pgfplotstablecreatecol[
    create col/expr={
        sqrt((pow(\thisrow{data1}-\thisrow{buildTemplate_mean},2) + pow(\thisrow{data2}-\thisrow{buildTemplate_mean},2) + pow(\thisrow{data3}-\thisrow{buildTemplate_mean},2) + pow(\thisrow{data4}-\thisrow{buildTemplate_mean},2) + pow(\thisrow{data5}-\thisrow{buildTemplate_mean},2) + pow(\thisrow{data6}-\thisrow{buildTemplate_mean},2) + pow(\thisrow{data7}-\thisrow{buildTemplate_mean},2) + pow(\thisrow{data8}-\thisrow{buildTemplate_mean},2) + pow(\thisrow{data9}-\thisrow{buildTemplate_mean},2) + pow(\thisrow{data10}-\thisrow{buildTemplate_mean},2))/10)
    }
]{buildTemplate_stddev}\buildTemplate
\pgfplotstablecreatecol[
    create col/expr={
        (\thisrow{data1} + \thisrow{data2} + \thisrow{data3} + \thisrow{data4} + \thisrow{data5} + \thisrow{data6} + \thisrow{data7} + \thisrow{data8} + \thisrow{data9} + \thisrow{data10})/10
    }
]{parseToJSON_mean}\parseToJSON

\pgfplotstablecreatecol[
    create col/expr={
        sqrt((pow(\thisrow{data1}-\thisrow{parseToJSON_mean},2) + pow(\thisrow{data2}-\thisrow{parseToJSON_mean},2) + pow(\thisrow{data3}-\thisrow{parseToJSON_mean},2) + pow(\thisrow{data4}-\thisrow{parseToJSON_mean},2) + pow(\thisrow{data5}-\thisrow{parseToJSON_mean},2) + pow(\thisrow{data6}-\thisrow{parseToJSON_mean},2) + pow(\thisrow{data7}-\thisrow{parseToJSON_mean},2) + pow(\thisrow{data8}-\thisrow{parseToJSON_mean},2) + pow(\thisrow{data9}-\thisrow{parseToJSON_mean},2) + pow(\thisrow{data10}-\thisrow{parseToJSON_mean},2))/10)
    }
]{parseToJSON_stddev}\parseToJSON
\pgfplotstablecreatecol[
    create col/expr={
        (\thisrow{data1} + \thisrow{data2} + \thisrow{data3} + \thisrow{data4} + \thisrow{data5} + \thisrow{data6} + \thisrow{data7} + \thisrow{data8} + \thisrow{data9} + \thisrow{data10})/10
    }
]{validate_mean}\validate

\pgfplotstablecreatecol[
    create col/expr={
        sqrt((pow(\thisrow{data1}-\thisrow{validate_mean},2) + pow(\thisrow{data2}-\thisrow{validate_mean},2) + pow(\thisrow{data3}-\thisrow{validate_mean},2) + pow(\thisrow{data4}-\thisrow{validate_mean},2) + pow(\thisrow{data5}-\thisrow{validate_mean},2) + pow(\thisrow{data6}-\thisrow{validate_mean},2) + pow(\thisrow{data7}-\thisrow{validate_mean},2) + pow(\thisrow{data8}-\thisrow{validate_mean},2) + pow(\thisrow{data9}-\thisrow{validate_mean},2) + pow(\thisrow{data10}-\thisrow{validate_mean},2))/10)
    }
]{validate_stddev}\validate

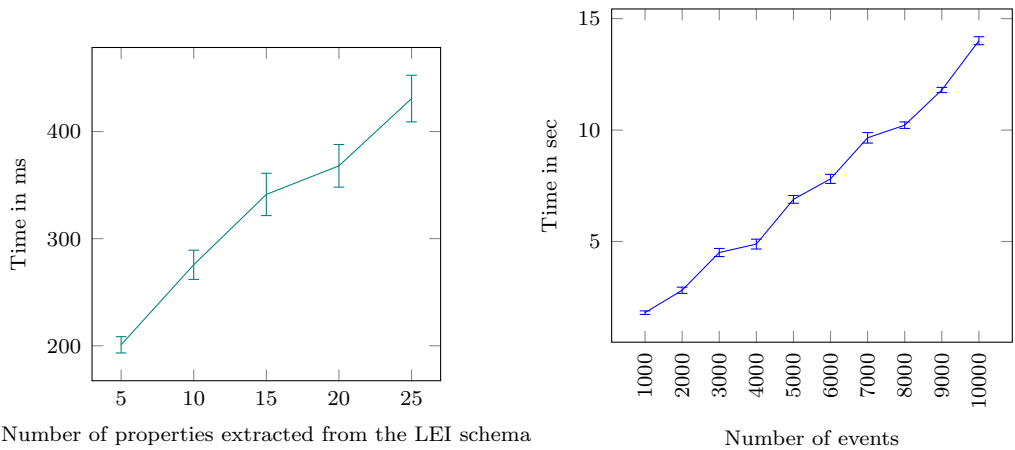
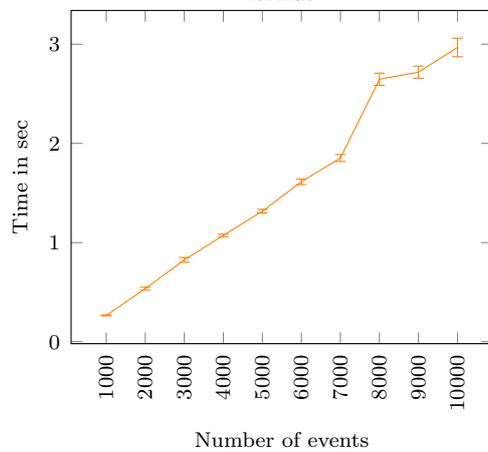
\begin{figure}[!htbp]
    \centering

    \begin{subfigure}{0.45\textwidth}
        \centering
\begin{tikzpicture}
\begin{axis}[
    xlabel={Number of properties extracted from the LEI schema},
    ylabel={Time in ms},
    width=\textwidth,
    height=6cm,
    grid=none,
    legend pos=south east,
    xtick=data,
    xticklabels from table={\buildTemplate}{properties},
    scaled y ticks=false,
    scaled x ticks=false
]
\addplot+[
    mark=none,
    teal,
    error bars/.cd,
    y explicit,
    y dir=both,
]
table [
    x=properties,
    y=buildTemplate_mean,
    y error=buildTemplate_stddev
] {\buildTemplate};

\end{axis}
\end{tikzpicture}
        \caption{Evaluating \function{buildTemplate} function -- the time required to create the spreadsheet template from the LEI schema}
        \label{fig:buildTemplate_function}
    \end{subfigure}\hfill
    \begin{subfigure}{0.5\textwidth}
        \centering
\begin{tikzpicture}
\begin{axis}[
    xlabel={Number of events},
    ylabel={Time in sec},
    width=\textwidth,
    height=6cm,
    grid=none,
    legend pos=south east,
            xtick={1000, 2000, 3000, 4000, 5000, 6000, 7000, 8000, 9000, 10000}, 
            xticklabels={1000, 2000, 3000, 4000, 5000, 6000, 7000, 8000, 9000, 10000}, 
    scaled y ticks=false,
    scaled x ticks=false,
    xticklabel style={rotate=90, anchor=east}, 
]
\addplot+[
    mark=none,
    blue,
    error bars/.cd,
    y explicit,
    y dir=both,
]
    table [
        x=events,
        y expr=\thisrow{parseToJSON_mean}/1000, 
        y error expr=\thisrow{parseToJSON_stddev}/1000  
    ] {\parseToJSON};

\end{axis}
\end{tikzpicture}
        \caption{Evaluating \function{parseToJSON} function -- the time required to convert the spreadsheet data into JSON format}
        \label{fig:parseToJSON}
    \end{subfigure}\hfill
    \begin{subfigure}{0.52\textwidth}
        \centering
\begin{tikzpicture}
\begin{axis}[
    xlabel={Number of events},
    ylabel={Time in sec},
    width=\textwidth,
    height=6cm,
    grid=none,
    legend pos=south east,
            xtick={1000, 2000, 3000, 4000, 5000, 6000, 7000, 8000, 9000, 10000}, 
            xticklabels={1000, 2000, 3000, 4000, 5000, 6000, 7000, 8000, 9000, 10000}, 
    scaled y ticks=false,
    scaled x ticks=false,
    xticklabel style={rotate=90, anchor=east}, 
]
\addplot+[
    mark=none,
    orange,
    error bars/.cd,
    y explicit,
    y dir=both,
]
    table [
        x=events,
        y expr=\thisrow{validate_mean}/1000, 
        y error expr=\thisrow{validate_stddev}/1000  
    ] {\validate};
\end{axis}
\end{tikzpicture}
        \caption{Evaluating \function{validate} function -- the time required to validate events (in JSON format) against the LEI schema}
        \label{fig:validate}
    \end{subfigure}
    \caption{Evaluation of the efficiency of the LEI2JSON tool}
    \label{fig:performance}
\end{figure}

\Cref{fig:performance} offers a detailed analysis and measures the time required to perform three key functions. For~\Cref{fig:buildTemplate_function}, the x-axis delineates the number of properties extracted from the schemas, ranging between 5 and 25 properties. On the contrary, the x-axis of~\Cref{fig:parseToJSON,fig:validate} corresponds to the count of events tested for a \textit{weight event} across eight columns of the spreadsheet, with values ranging from 1,000 to 10,000 events. The universally applied y-axis on all diagrams measures the time in milliseconds for \function{buildTemplate} and in seconds for \function{parseToJSON} and \function{validate}.

Upon analysis, distinct efficiency trends become evident. The functions \function{buildTemplate}, \function{parseToJSON}, and \function{validate} display a linear increase in operational time as the number of properties or events extracted increases. This implies a direct proportionality between their operational time and the respective count. The results indicate that LEI2JSON maintains a consistent efficiency rate for each functionality, regardless of the size or complexity of the data. This suggests that LEI2JSON can handle large and diverse datasets without compromising performance or quality. However, the results also reveal variations in efficiency rates among different functionalities, depending on their time consumption per unit of data. This highlights the potential for optimisation in LEI2JSON's functions and processes to reduce execution time and resource consumption.

\section{Impact}\label{sec:impact}

In this study, we introduce LEI2JSON, an application designed to streamline the conversion of spreadsheet data into a standardised JSON format for livestock events. Integrated into Google Sheets through a user-friendly HTML sidebar, LEI2JSON caters to a diverse user base, ranging from livestock producers to researchers in various agricultural sectors.

This tool transforms livestock event data into JSON messages while strictly adhering to the LEI schema and ensuring real-time data validation. Users can easily copy, download, or save the generated JSON text to Google Drive.

The significance of this conversion lies in its potential to advance research, especially in the realms of data management and event-based messaging within the red meat industry. LEI2JSON's key features include independence from column ordering during JSON generation, real-time data validation, and automated spreadsheet template creation based on LEI schema specifications.

Anticipated to profoundly impact the market, LEI2JSON is poised to improve regulatory efficiency and productivity for red meat producers that embrace the JSON message. By streamlining data collection and organisation for livestock events, this tool empowers producers to improve the value of their farming enterprises.

\section{Conclusion}\label{sec:conclusion}

LEI2JSON is a Google Sheets add-on that plays a crucial role in standardising livestock data in JSON format according to the Livestock Event Information (LEI) JSON schema. This standardisation empowers stakeholders to make informed decisions and increase profitability. The seamless integration of HTML with Google Apps Script ensures efficient execution of key functions, including creating a spreadsheet template, generating JSON, performing validation, and offering versatile sharing options.

Looking ahead, there is exciting potential for expansion to cover more events and agricultural sectors, along with improvements in the user interface. We argue that LEI2JSON has the potential to make a substantial impact on the livestock industry.

\section*{Funding sources}
This article was supported by funding from Food Agility CRC Ltd, funded under the Commonwealth Government CRC Program. The CRC Program supports industry-led collaborations between industry, researchers, and the community. This manuscript was also funded by the Gulbali Institute Accelerated Publication Scheme (GAPS).


\section*{Acknowledgements}
The authors thank David Swain and Will Swain from TerraCipher for their guidance and assistance throughout the article.

  \bibliographystyle{elsarticle-num-names} 
  \bibliography{refs}
  
\end{document}